%% file: 9408314.tex
\documentstyle[12pt]{article}

\newcommand{\reseteqnum}{\setcounter{equation}{0}}

\newcommand{\rbox}[1]{\vbox{\hrule height.8pt%
                \hbox{\vrule width.8pt\kern5pt
                \vbox{\kern5pt\hbox{#1}\kern5pt}\kern5pt
                \vrule width.8pt}
                \hrule height.8pt}}

\begin{document}
\baselineskip=18pt plus 0.2pt minus 0.1pt
\renewcommand{\thepage}{}


\begin{titlepage}
\title{
\vspace{-5ex}
\hfill
\parbox{4cm}{\normalsize KUNS-1287\\HE(TH)\ 94/11\\
hep-ph/9408314}\\
\vspace{2ex}
Hierarchical Mass Matrices
       \break in a Minimal {\it SO}(10) Grand Unification
\uppercase\expandafter{\romannumeral2}
\vspace{2ex}}

\setcounter{footnote}{0}

\author{Masako {\sc Bando}
\thanks{e-mail address: {\tt mband@jpnyitp.bitnet}}
\thanks{Supported in part by Grant-in-Aid for Scientific Research from
Ministry of Education, Science, and Culture \hspace*{2em}(\#06640416).}\\
{\small\em Physics Division, Aichi University}\\
{\small\em Aichi 470-02, Japan}\vspace{2ex}\\
{\sc Izawa} Ken-Iti
\thanks{JSPS Research Fellow.}\\
{\small\em Department of Physics, University of Tokyo}\\
{\small\em Tokyo 113, Japan}\vspace{1.5ex}\\
and\vspace{1.5ex}\\
 Tomohiko {\sc Takahashi}
\thanks{e-mail address: {\tt tomo@gauge.scphys.kyoto-u.ac.jp}}
\thanks{JSPS Research Fellow.}\\
{\small\em Department of Physics, Kyoto University}\\
{\small\em Kyoto 606-01, Japan}}

\date{August, 1994}
\maketitle

\vspace{5ex}
\begin{abstract}
\normalsize
We continue to
investigate the minimal $SO(10)\times U(1)_H$ model which we
proposed recently. Renormalization group
analysis of the model
results in natural predictions
of quark-lepton masses and Kobayashi-Maskawa matrix
along with neutrino mixings adequate for solar neutrino oscillation.
\end{abstract}
\end{titlepage}

\newpage
\renewcommand{\thepage}{\arabic{page}}
\setcounter{page}{2}
\section{Introduction}
\reseteqnum

Grand unification has been a fascinating idea for two decades.
Among the possible gauge groups, $SO(10)$ is the smallest
candidate that can incorporate the observed fermions of one generation
into an irreducible multiplet:
it attains matter unification
for a single generation of quarks and leptons.
However, $SO(10)$ grand unification by itself provides no natural place
for triplicity of generations,
to say nothing of the hierarchical structure
of mass matrices.
For instance,
it gives no explanation on the fact that the mass of top quark
is more than $10^6$ times that of electron.

In a previous paper
\cite{BIT},
we constructed an $SO(10)\times U(1)_H$ model
as an attempt to place generation structure
on a plausible position in grand unification.
We introduced a {\em minimal} Higgs
content to break the gauge symmetry
without any
additional scalars.
The horizontal Peccei-Quinn symmetry $U(1)_H$
distinguishes three generations of fermions
to impose Georgi-Jarlskog(GJ) relations
\cite{GJ}
at the unification scale $M_U$.
It also makes
Yukawa couplings of order one
be able to realize hierarchical mass matrices
with the aid of remnant effects
of certain irrelevant
terms suppressed by the Planck scale $M_P$
as the cut-off scale in the theory
(For more details, see
ref.\cite{BIT}).

The model has three factors which determine the values of mass
matrices in combination with order-one Yukawa couplings
at the unification scale.
The first one is the hierarchy factor $\varepsilon = M_U/M_P$
brought about by the remnant effects mentioned above.
The second one comes from the mixing of Higgs bosons
which is represented by
parameters $\alpha$, $\beta$, $\gamma$, $\delta$ defined later.
The third is the running of Yukawa couplings
according to renormalization group\,(RG) equations.

In this paper, we proceed to
analyze mainly the third factor to compare
predictions of our model with experimental estimates.
Numerical analysis of RG equations
results in natural predictions
of quark-lepton masses and Kobayashi-Maskawa(KM) matrix
along with neutrino mixings adequate for solar neutrino oscillation
\cite{FukuYana}.

The paper is organized as follows: We first recapitulate the setup of our
model in section 2 to derive RG equations in section 3.
We restrict ourselves to dealing with one-loop RG equations
and ignore threshold corrections throughout the paper.
Section 4 makes exposition of numerical
results obtained by RG analysis.
Section 5 concludes the paper.
Some definitions are given in Appendix A
and analytical consideration on quark mass matrices
is made in Appendix B.

\section{Yukawa Interactions}
\reseteqnum

In this section, we present
effective Yukawa interactions
under the chain of symmetry breaking considered in
ref.\cite{BIT}.
There are three scales of symmetry breaking postulated
(with the help of fine-tuning) in the model:
the unification scale $M_U$, the intermediate scale $M_I$,
and the weak scale $M_W$.

We introduced Higgs fields $\Phi (210, -8)$,
$\bar{\Delta} (\overline{126}, -10)$, and $H(10, -2)$
transforming under $SO(10)\times U(1)_H$.
The field $\Phi (210, -8)$
develops a VEV of order $M_U$,
which breaks
$SO(10)\times U(1)_H$ into $G_{224}\times Z_8$,
where $G_{224}$ denotes the Pati-Salam group
$SU(2)_L\times SU(2)_R\times SU(4)_C$.
In particular, D-parity oddness of $\Phi (210, -8)$ causes
violation of the left-right symmetry.

We assumed that there remain $H(2, 2, 1)$,
$\bar{\Delta} (2, 2, 15)$, and
$\bar{\Delta} (1, 3, \overline{10})$
on the Pati-Salam stage between the scales $M_U$ and $M_I$,
the other Higgs fields with masses of order $M_U$
decoupling from the system.
This is a modification of
the minimal fine-tuning,
which claims non-decoupling of only
$H(2, 2, 1)$ and $\bar{\Delta} (1, 3, \overline{10})$.
The additional field $\bar{\Delta} (2, 2, 15)$
develops an induced VEV
\cite{Laz}
of a considerable size
to contribute fermion mass matrices
\cite{BIT}.

The effective Yukawa interactions
on this stage are given by
\begin{eqnarray}
 {\cal L}_{1} &=& Y_1 \psi^T_L \phi_{1} \psi_R
  + Y_2 \psi^T_L
  \widetilde{\phi}_{1} \psi_R \nonumber\\
  &&+ Y_3 \psi_L^T \phi_{15} \psi_R
  + Y_4 \psi_L^T \widetilde{\phi}_{15}
  \psi_R + Y_5 \psi_R^T \phi_{10} \psi_R + h.c.,
 \label{eq:yukawa1}
\end{eqnarray}
where
$Y$'s are Yukawa couplings
and summation over suppressed generation indices should be understood.
Fermions $\psi_L$ and $\psi_R$ denote
$(2,1,4)$ and $(1,2,{\overline{4}})$ representations,
respectively.
Scalars
$\phi_1$, $\phi_{15}$, and $\phi_{10}$ correspond to
$H(2, 2, 1)$,
$\bar{\Delta} (2, 2, 15)$, and
$\bar{\Delta} (1, 3, \overline{10})$
in this turn
(See Appendix A).

Yukawa couplings at the unification scale $M_U$
provide a boundary condition
\begin{eqnarray}
 Y_1 = \left(
  \begin{array}{ccc}
  -\varepsilon^{\,2} y_{11} & 0 & \varepsilon y_{13}\\
  0 & 0 & 0 \\
  -\varepsilon y_{13} & 0 & y_{33}\\
  \end{array}
  \right),
  \hspace{36pt}
 Y_2 = \left(
  \begin{array}{ccc}
  0 & \varepsilon^{\,2} y_{12} & 0 \\
  \varepsilon^{\,2} y_{12} & 0 & -\varepsilon y_{23}\\
  0 & \varepsilon y_{23} & 0
  \end{array}
  \right),
 \nonumber
\end{eqnarray}
\begin{eqnarray}
 Y_3 = \left(
  \begin{array}{ccc}
  0 & 0 & z_{13} \\
  0 & z_{22} & 0 \\
  z_{13} & 0 & 0
  \end{array}
  \right),
  \hspace{36pt}
 Y_4 = 0,
  \hspace{36pt}
 Y_5 = \frac{1}{\sqrt{2}}\left(
  \begin{array}{ccc}
  \varepsilon z_{11} & 0 & z_{13} \\
  0 & z_{22} & 0 \\
  z_{13} & 0 & \varepsilon z_{33}
  \end{array}
  \right),
\end{eqnarray}
where Y's have been regarded as matrices in generation space.
The small factor $\varepsilon=M_U/M_P$
stems from the remnant effects and $y$'s and $z$'s
are input parameters of order one, which we expect to be determined
by more fundamental theory presumably including gravitation.
GJ relations can be obtained from the above texture
as shown in
ref.\cite{BIT}
(See also Appendix B).

Below the intermediate scale $M_I$, the model effectively
coincides with the standard model, and
the effective Yukawa interactions are given by
\begin{eqnarray}
 {\cal L}_2&=&Y^\dagger_d\ ({\overline Q}\,\phi)\,d_R
  +Y^\dagger_u\,({\overline Q}
  \,\widetilde{\phi})\,u_R
  + Y^\dagger_e\,({\overline L}\,\phi)\,e_R
  +Y^\dagger_\nu\,({\overline L}\,\widetilde{\phi})\,\nu_R + h.c.,
 \label{eq:yukawa2}
\end{eqnarray}
where $\phi$, $Q$, and $L$ denote the standard Higgs, quark,
and lepton doublets, respectively.
Although this Lagrangian contains
Dirac mass terms for neutrinos
$Y^\dagger_\nu({\overline L}\widetilde{\phi})\nu_R$,
they can be approximately neglected
\cite{NEU}
below $M_I$
due to order-$M_I$ Majorana masses of right-handed neutrinos $\nu_R$,
which originates from the term $Y_5 \psi_R^T \phi_{10} \psi_R$
in eq.(\ref{eq:yukawa1}).

The standard Higgs doublet $\phi$ is a linear combination
\cite{GN}
of four doublets contained in $H(2, 2, 1)$
and $\bar{\Delta}(2, 2, 15)$:
\begin{eqnarray}
 \phi = \alpha H_{\frac{1}{2}}
      + \beta {\widetilde H}_{-\frac{1}{2}}
      + \gamma \bar{\Delta}_{\frac{1}{2}}
      + \delta \widetilde{\bar{\Delta}}_{-\frac{1}{2}}
 \label{eq:standardhiggs}
\end{eqnarray}
with a normalization condition
\begin{eqnarray}
 \alpha^{\,2}+\beta^{\,2}+\gamma^{\,2}+\delta^{\,2}=1,
 \label{eq:constraint}
\end{eqnarray}
where subscripts $\pm \frac{1}{2}$ indicate hypercharges of the fields.
Hence the following relations hold at
the intermediate scale $M_I$:
\begin{eqnarray}
 Y_u&=&\alpha Y_1 +\beta Y_2 +\frac{1}{2\sqrt{3}}\,\gamma Y_3
  +\frac{1}{2\sqrt{3}}\,\delta Y_4, \nonumber\\
 Y_d&=&\alpha Y_2 +\beta Y_1 +\frac{1}{2\sqrt{3}}\,\gamma Y_4
  +\frac{1}{2\sqrt{3}}\,\delta Y_3, \nonumber\\
 Y_\nu&=&\alpha Y_1 +\beta Y_2 +\frac{-3}{2\sqrt{3}}\,\gamma Y_3
  +\frac{-3}{2\sqrt{3}}\,\delta Y_4, \nonumber\\
 Y_e&=&\alpha Y_2 +\beta Y_1 +\frac{-3}{2\sqrt{3}}\,\gamma Y_4
  +\frac{-3}{2\sqrt{3}}\,\delta Y_3,
\end{eqnarray}
where the coefficients $\frac{1}{2\sqrt{3}}$ and $\frac{-3}{2\sqrt{3}}$
result from the normalization of $\phi_{15}$ defined in the Appendix A.
The parameters  $\alpha$, $\beta$, $\gamma$,
$\delta$ are written in terms of the VEVs
defined in
ref.\cite{BIT}
as follows:
\begin{eqnarray}
 \alpha = \frac{v_t}{v_W},
  \hspace{1em}
 \beta = \frac{v_b}{v_W},
  \hspace{1em}
 \gamma = 2\sqrt{3}\frac{w_c^*}{v_W},
  \hspace{1em}
 \delta = 2\sqrt{3}\frac{w_s^*}{v_W},
\end{eqnarray}
where $v_W$ denote the VEV of the standard Higgs doublet.
Note that the VEVs $w_c^*$ and $w_s^*$
are of order $\varepsilon$ relative to $v_W$
under the assumption in
ref.\cite{BIT}
that they are induced in the Higgs $\phi_{15}$
with mass of order $M_I$.
Hence $\gamma$ and $\delta$
take values of order $\varepsilon$.
This makes it natural to define order-one quantity
$\gamma'$ and $\delta'$ in terms of
\begin{eqnarray}
 \gamma = \gamma' \varepsilon, \hspace{2em}
 \delta = \delta' \varepsilon.  \label{eq:gamdel}
\end{eqnarray}

It should be emphasized that only two parameters $\varepsilon$
and $\beta$ are small to realize the hierarchical structure
of mass matrices (See Appendix B).
The other parameters $y$'s, $z$'s, $\gamma'$, and $\delta'$
are of order one, which merely affect detailed numerical values
of masses and mixings without altering their orders of magnitude.

\section{Renormalization Group Equations}
\reseteqnum

In this section, we show one-loop RG equations
for gauge and Yukawa
\cite{Machacek}
couplings ($g$'s and $Y$'s) in the effective theories
described in the previous section.

We first give RG equations for the $G_{224}$ theory
on the stage between the unification and intermediate scales:
\begin{eqnarray}
 \frac{d\omega_i}{dt}=-\frac{1}{2\pi}b_i; \hspace{24pt}
  b_i=\left(2, \frac{26}{3}, -\frac{7}{3}\right),
  \hspace{24pt}i=2_L,\ 2_R,\ 4_C,
 \label{eq:gauge1}
\end{eqnarray}
where $\omega_i = \frac{4\pi}{g_i^2}$,
$t={\rm ln}\mu$, and $\mu$ denotes a renormalization point
in the $\overline{{\rm MS}}$ scheme; and
\begin{eqnarray}
 16\pi^{\,2}\frac{dY_1}{dt}&=&Y_1 \beta_L+\beta_R^\dagger Y_1+Y_1 \beta_1
 +\beta_{v1}+Y_1 \beta_g, \nonumber\\
 16\pi^{\,2}\frac{dY_2}{dt}&=&Y_2 \beta_L+\beta_R^\dagger Y_2+Y_2 \beta_1
 +\beta_{v2}+Y_2 \beta_g, \nonumber\\
 16\pi^{\,2}\frac{dY_3}{dt}&=&Y_3 \beta_L+\beta_R^\dagger Y_3+Y_3 \beta_{15}
 +\beta_{v3}+Y_3 \beta_g, \nonumber\\
 16\pi^{\,2}\frac{dY_4}{dt}&=&Y_4 \beta_L+\beta_R^\dagger Y_4+Y_4 \beta_{15}
 +\beta_{v4}+Y_4 \beta_g, \nonumber\\
 16\pi^{\,2}\frac{dY_5}{dt}&=&Y_5 \beta_R+\beta_R^\dagger Y_5+Y_5
 \beta_{10}+Y_5 \beta_g',
\end{eqnarray}
where $\beta_L$, $\beta_R$ and $\beta_{1,\ 15,\ 10}$ correspond to
contributions from wave-function renormalization of $\psi_L$,
$\psi_R$, and $\phi_{1,\ 15,\ 10}$, respectively;
$\beta_v$'s and $\beta_g$, $\beta_g'$
correspond to contributions from vertex renormalization and gauge
couplings:
\begin{eqnarray}
 \beta_L&=&Y_1^\dagger Y_1+Y_2^\dagger Y_2 +\frac{15}{4}(Y_3^\dagger Y_3
  +Y_4^\dagger Y_4), \nonumber\\
 \beta_R&=&Y_1^\dagger Y_1+Y_2^\dagger Y_2+\frac{15}{4}(Y_3^\dagger
  Y_3+Y_4^\dagger Y_4+Y_5^\dagger Y_5), \nonumber\\
 \beta_1&=&4\ {\rm tr}(Y_1^\dagger Y_1+Y_2^\dagger Y_2), \nonumber\\
 \beta_{15}&=&{\rm tr}(Y_3^\dagger Y_3+Y_4^\dagger Y_4), \nonumber\\
 \beta_{10}&=&{\rm tr}(Y_5^\dagger Y_5), \nonumber\\
 \beta_{v1}&=&-2Y_1Y_2^\dagger Y_2-2Y_2Y_2^\dagger Y_1
  -\frac{15}{2}Y_3Y_2^\dagger Y_4-\frac{15}{2}Y_4Y_2^\dagger Y_3, \nonumber\\
 \beta_{v2}&=&-2Y_2Y_1^\dagger Y_1-2Y_1Y_1^\dagger Y_2
  -\frac{15}{2}Y_3Y_1^\dagger Y_4-\frac{15}{2}Y_4Y_1^\dagger Y_3, \nonumber\\
 \beta_{v3}&=&\frac{1}{2}Y_3Y_4^\dagger Y_4
  +\frac{1}{2}Y_4Y_4^\dagger Y_3-2Y_1Y_4^\dagger Y_2-2Y_2Y_4^\dagger Y_1,
  \nonumber\\
 \beta_{v4}&=&\frac{1}{2}Y_3Y_3^\dagger Y_4
  +\frac{1}{2}Y_4Y_3^\dagger Y_3-2Y_1Y_3^\dagger Y_2-2Y_2Y_3^\dagger Y_1,
  \nonumber\\
 \beta_g&=&\frac{9}{4}g_L^{\,2}+\frac{9}{4}g_R^{\,2}+\frac{15}{4}g_{4C}^{\,2},
  \nonumber\\
\beta_g'&=&\frac{9}{2}g_R^{\,2}+\frac{15}{4}g_{4C}^{\,2}.
\end{eqnarray}

Let us turn to the energy region below the intermediate scale.
RG equations on this stage are those for the standard model:
\begin{eqnarray}
 \frac{d\omega_i}{dt}=-\frac{1}{2\pi}b_i; \hspace{24pt}
  b_i=\left(\frac{41}{10},-\frac{19}{6},-7\right),
  \hspace{24pt}i=1_Y,\ 2_L,\ 3_C;
 \label{eq:gauge2}
\end{eqnarray}
\begin{eqnarray}
 16\pi^{\,2}\frac{dY_u}{dt}&=&Y_u\left[\frac{3}{2}(Y^\dagger_u
  Y_u-Y^\dagger_d Y_d)+{\rm tr}(3Y^\dagger_u Y_u+3Y^\dagger_d
  Y_d+Y^\dagger_e Y_e)
  -\left(\frac{17}{20}g_1^{\,2}
  +\frac{9}{4}g_2^{\,2}+8g_3^{\,2}\right)\right], \nonumber\\
 16\pi^{\,2}\frac{dY_d}{dt}&=&Y_d\left[\frac{3}{2}(Y^\dagger_d
  Y_d-Y^\dagger_u Y_u)+{\rm tr}(3Y^\dagger_u Y_u+3Y^\dagger_d
  Y_d+Y^\dagger_e Y_e)
  -\left(\frac{1}{4}g_1^{\,2}+\frac{9}{4}g_2^{\,2}
  +8g_3^{\,2}\right)\right], \nonumber\\
 16\pi^{\,2}\frac{dY_e}{dt}&=&Y_e\left[\frac{3}{2}Y^\dagger_e
  Y_e+{\rm tr}(3Y^\dagger_u Y_u+3Y^\dagger_d
  Y_d+Y^\dagger_e Y_e)
  -\frac{9}{4}(g_1^{\,2}+g_2^{\,2})\right].
 \label{eq:rgesm}
\end{eqnarray}
For a recent analysis of them, see
Ref.\cite{Arason}.

\section{Numerical Results}
\reseteqnum

We proceed to consider numerical solutions to the RG equations
listed in the previous section.
The running of the gauge couplings are
independent of Yukawa couplings to the
extent of one-loop analysis. Thus we can first obtain the
unification scale $M_U$ and the intermediate scale $M_I$
\cite{Deshpa}
by means of
RG equations for gauge couplings (\ref{eq:gauge1}) and (\ref{eq:gauge2})
with their experimental values below the weak scale
\cite{PDG}
as a boundary condition:
\begin{eqnarray}
 M_U \simeq 10^{\,16.7}\,{\rm GeV}, \hspace{1em}
 M_I \simeq 10^{\,11.2}\,{\rm GeV}.
\end{eqnarray}
The unified gauge coupling at $M_U$ comes out to be
$g_U \simeq 0.585$. We think of this as a typical value of order one,
since gauge coupling seems fundamental in view of its geometrical origin.
Yukawa couplings at the unification scale are to be compared with this value
as the standard one.
(Conversely, one might also regard this value as an experimental evidence
of coupling constants being of order one.)

Now that we have obtained running gauge couplings,
we turn to analyzing Yukawa sector.
We assume CP conservation in the following analysis.
In particular, KM phase is set to zero.
Thus the case of large KM phase are excluded from the analysis
in this paper,
though small phase may be taken into account perturbatively
and seems not to affect the results considerably.

The procedure we pursue is as follows:
To begin with, we choose appropriate values
for the input parameters
$\varepsilon$, $\beta$,
$y$'s, $z$'s, $\gamma'$, and $\delta'$
partly with the help of trial and error
(See Appendix B).
Note that $y$'s, $z$'s, $\gamma'$, and $\delta'$
must be of order one.
Then we make Yukawa couplings evolve
from the unification scale $M_U$ down to the weak scale
$M_W \simeq 174{\rm GeV}$ by solving
RG equations numerically. Finally the resultant Yukawa
matrices at the weak scale
are diagonalized to yield fermion masses and mixings.
Neutrino masses and mixings
are derived by means of sea-saw approximation
\cite{FukuYana}
from the values of Yukawa couplings
at the intermediate scale $M_I$, where the right-handed neutrinos
are supposed to decouple.
We compare the results with experimental estimates of
running masses and mixings at the weak scale
\cite{Koide}.

Let us exhibit a numerical sample which provides a realistic pattern
of mass matrices at the weak scale:
input parameters in table \ref{tbl: input} lead to the results in
tables \ref{tbl: outmass} and \ref{tbl: outkm}.
The masses of neutrinos are given by
\begin{eqnarray}
 ({\rm m_{\nu_e},\  m_{\nu_\mu},\  m_{\nu_\tau}})
 \sim (4.1 \times 10^{-11},\ 5.9 \times 10^{-5},\ 3.0 \times 10^{-2})
 \times \frac{v_W^{\,2}}{v_I},
\end{eqnarray}
where $v_I$ denotes the VEV developed by the Higgs $\phi_{10}$.
The values $v_W \simeq 174{\rm GeV}$ and $v_I \sim 10^{11.2}{\rm GeV}$
predict
\begin{eqnarray}
 m_{\nu_\mu}\sim 10^{-3} {\rm eV},
\end{eqnarray}
and a negligible value of $m_{\nu_e}$ relative to $m_{\nu_\mu}$.
This is consistent with the small-angle MSW solution
to the solar neutrino problem
\cite{FukuYana},
which is implemented by
\begin{eqnarray}
 \Delta m_{e\mu} &=& (m_{\nu_\mu}^{\,2}-m_{\nu_e}^{\,2})^{\,1/2}
  \sim 10^{-3} {\rm eV}, \nonumber\\
 {\rm sin}\theta_{e\mu} &\simeq& 0.03 - 0.06.
\end{eqnarray}

The above example
shows that qualitative agreement is achieved between
predictions of the model and experimental estimates.
In particular, the hierarchical structure of mass matrices
was shown to be
indeed realized in terms of Yukawa couplings
exclusively of order one.

%
%
\begin{table}[p]
\begin{eqnarray}
 Y_1=\left(
  \begin{array}{ccc}
   0.3\times \varepsilon^{\,2} & 0 & 0.3\times \varepsilon\\
   0 & 0 & 0 \\
   -0.3\times \varepsilon & 0 & 0.5\\
  \end{array}
  \right),
  \hspace{18pt}
 Y_2=\left(
  \begin{array}{ccc}
   0 & 3.25\times \varepsilon^{\,2} & 0 \\
   3.25\times \varepsilon^{\,2} & 0 & 0.16\times \varepsilon \\
   0 & -0.16\times \varepsilon & 0
  \end{array}
  \right), \nonumber
\end{eqnarray}\vspace{-6.5ex}

\begin{eqnarray}
 Y_5=\left(
  \begin{array}{ccc}
   0.5 \times \varepsilon & 0 & -0.145 \\
   0 & 0.65 & 0 \\
   -0.145 & 0 & 0.5 \times \varepsilon
  \end{array}
  \right), \nonumber
\end{eqnarray}\vspace{-5ex}

\begin{eqnarray}
 \varepsilon=\frac{1}{250},
 \hspace{2em}
 \beta=\frac{1}{53}, \hspace{2em}
 \gamma'=\frac{25}{9}, \hspace{2em}\delta'=\frac{1}{3}. \nonumber
\end{eqnarray}

\caption{Sample input parameters at $\mu = 10^{16.7}{\rm GeV}$.}
\label{tbl: input}
\end{table}

%
%
\begin{table}[p]
\begin{tabular}{|c|c|c||c|c|c|}
 \hline
 & \makebox[6em]{Prediction} & \makebox[6em]{Experiment} & &
 \makebox[6em]{Prediction} & \makebox[6em]{Experiment} \\
 \hline
 \hline
 $m_u$ & $2.5\times 10^{-3}$ & $2.4\times 10^{-3}$&
 $m_d$ & $3.3\times 10^{-3}$ & $4.2\times 10^{-3}$ \\
 \hline
 $m_c$ & 0.61 & 0.61 & $m_s$ & 0.087 & 0.085\\
 \hline
 $m_t$ & 160 & & $m_b$ & 3.7 & 2.9\\
 \hline
 \hline
 & \makebox[6em]{Prediction} & \makebox[6em]{Experiment} & &
 \makebox[6em]{Prediction} & \makebox[6em]{Experiment} \\
 \hline
 \hline
 $m_e$ & $5\times 10^{-4}$ & $5\times 10^{-4}$ &
 $m_{\nu_e}$ & $\sim 4 \times 10^{-19}$ &    \\
 \hline
 $m_\mu $ & 0.1 & 0.1 &
 $m_{\nu_\mu}$ & $\sim 6 \times 10^{-13}$ &  \\
 \hline
 $m_\tau $ & 1.7 & 1.7 &
 $m_{\nu_\tau}$ & $\sim 3 \times 10^{-10}$ &  \\
 \hline
\end{tabular}\vspace{1em}

\caption{Running quark and lepton masses (GeV) at $\mu = 174{\rm GeV}$.}
\label{tbl: outmass}
\end{table}

%
%
\begin{table}[p]
\begin{eqnarray}
 V_{\rm quark}&=&\left(
  \begin{array}{ccc}
   -0.98 & 0.22 & -0.0063 \\
   -0.22 & -0.98 & 0.052 \\
   -0.005 & -0.052 & -1
  \end{array}
  \right)\nonumber\\
 V_{\rm lepton}&=&\left(
  \begin{array}{ccc}
   \ \ 1 & -0.058 & \ -0.017 \\
   \ \ 0.059 & 1 & \ 0.062 \\
   \ \ 0.014 & \ -0.063 & 1
  \end{array}
  \right)\nonumber
\end{eqnarray}

\caption{Predictions of quark and lepton mixing
matrices at $\mu = 174{\rm GeV}$.}
\label{tbl: outkm}
\end{table}

So far so good.
However, we cannot help complaining
about the prediction of bottom mass.
It has tendency to come out larger
\cite{Keith}
in this model than the experimental estimate
(provided the tau mass is fitted.)
This defect originates from GJ relation
$m_b \simeq m_{\tau}$ at the unification scale
and is scarcely dependent on the parameters
we can choose to get realistic predictions.
Taking this issue seriously,
we might need some modifications of the present model.
Even in such circumstances, we hope that the qualitative
features of the model survive to naturally achieve the mass-matrix hierarchy.

\section{Concluding Remarks}
\reseteqnum

In the preceding sections,
we have investigated the running
of Yukawa couplings below the unification scale
in our model of grand unification
\cite{BIT}.
The remnant effects of orders
$\varepsilon$ and $\varepsilon^2$
play crucial roles to make realistic predictions
in the model.
Without the remnants, the model would
possess a parity symmetry for the second generation
$\psi_2 \rightarrow -\psi_2$,
which completely forbids mixings between the second and the other
generations:
that is,
in the renormalizable setting,
radiative corrections could not
generate necessary operators corresponding to the remnants.

Finally
let us see the behavior of couplings
above the unification scale.
The Yukawa interactions are given by
\begin{eqnarray}
 {\cal L}_0=\frac{1}{4}\,Y_{10} \psi^T B \gamma_\mu H_\mu \psi
            + \frac{1}{4\cdot 5!}\,Y_{126}  \psi^T B \gamma_{\mu_1 \mu_2
            \cdots \mu_5} \bar{\Delta}_{\mu_1 \mu_2 \cdots \mu_5}
            \psi + h.c.,
\end{eqnarray}
where higher-dimensional terms are ignored.
Greek indices are $SO(10)$ vector ones,
$\gamma_\mu$ yields 32-dimensional
representation of Clifford algebra
$\left\{\,\gamma_\mu\,,\,\gamma_\nu\,\right\}=2\,\delta_{\mu\,\nu}$,
and $B$ denotes charge conjugation for $SO(10)$ spinors:
$B=\prod_{\mu:\,{\rm odd}}\,\gamma_\mu$.
126 Higgs
$\bar{\Delta}$ is self-dual antisymmetric tensor,
which satisfy
\begin{eqnarray}
 \bar{\Delta}_{\mu_1 \mu_2 \cdots \mu_5}
  =\frac{i}{5!}\,\epsilon_{\mu_1 \mu_2
  \cdots \mu_5 \mu_6 \mu_7 \cdots \mu_{10}}\,
  \bar{\Delta}_{\mu_6 \mu_7
  \cdots \mu_{10}},
\end{eqnarray}
where $\epsilon_{\mu_1 \mu_2 \cdots \mu_{10}}$ denotes the
invariant antisymmetric tensor.

The one-loop RG equation for the unified gauge coupling $g$ is given by
\begin{eqnarray}
 \frac{d\omega}{dt} = -\frac{1}{2\pi} \frac{16}{3},
\end{eqnarray}
which shows that the theory is asymptotically non-free.
Note that the absolute value of the beta function
is so small that the coupling stays within perturbative range
up to the Planck scale.
The running of effective
gauge couplings are
exhibited in figure \ref{fig: gauge}.

The Yukawa couplings above the unification scale obey
approximate RG equations
\begin{eqnarray}
 16\pi^2 \frac{dY_{10}}{dt} &=&
                        \frac{10}{16}\, Y_{10}Y_{10}^\dagger Y_{10}
                        + \frac{63}{8}\,Y_{126} Y_{126}^\dagger Y_{10}
                        + \frac{63}{8}\,Y_{10} Y_{126}^\dagger Y_{126}
                        \nonumber\\
                    && + Y_{10} {\rm tr}(Y_{10}Y_{10}^\dagger)
                        -24\, g^{\,2} Y_{10} ,
\end{eqnarray}
\begin{eqnarray}
 16\pi^2 \frac{dY_{126}}{dt} &=&
                         \frac{63}{4}\,Y_{126} Y_{126}^\dagger Y_{126}
                         + \frac{5}{16}\,Y_{10} Y_{10}^\dagger Y_{126}
                         + \frac{5}{16}\,Y_{126} Y_{10}^\dagger Y_{10}
                         \nonumber\\
                     && + 2\,Y_{126} {\rm tr}(Y_{126} Y_{126}^\dagger)
                         -24\,g^{\,2} Y_{10}
\end{eqnarray}
At the unification scale,
they satisfy
\begin{eqnarray}
 (Y_{10})_{\,33}=(Y_1)_{\,33}, \hspace{2em}2\,Y_{126}=Y_3.
\end{eqnarray}

The flow of Yukawa coupling $(Y_{10})_{33}$ with respect to
the unified gauge coupling $g$
is shown in figure \ref{fig: bitop}.
The Plank scale corresponds to
$g \simeq 0.62$ and the GUT scale to
$g \simeq 0.585$.
Thus the Yukawa coupling $y_{33}$
in the numerical sample in the previous section
is found to be of order one even at the Planck scale.
This is consistent with the perturbative treatment above
and the general philosophy of effective field theory
that coupling constants are of order one at the cut-off scale.

In fact, figure \ref{fig: bitop}
suggests that $(Y_{10})_{33}$ is almost always of order one at the GUT scale
whatever it is at the Plank scale.
One can even consider the case where the Yukawa
coupling blows up at the Plank scale,
which might indicate the presence of some dynamical phenomenon
out of perturbative picture.
\vspace{3ex}

\begin{figure}[h]
\input{gauge}
\caption{Running gauge couplings;}
\centerline{$\omega = 4\pi/g^2$ and $\mu$ is a renormalization point
in GeV unit.}
\label{fig: gauge}
\end{figure}

\newpage

\begin{figure}[t]
\input{bitop}
\caption{Running Yukawa coupling above $M_U$;}
\centerline{$g$ denotes the unified gauge coupling.}
\label{fig: bitop}
\end{figure}
\vspace{8ex}


\centerline{\large\bf Acknowledgments}
We would like to thank M.~Harada and T.~Kugo for
helpful discussions. We also acknowledge
correspondence with Y.~Koide
on running quark masses below the weak scale.
I.~K.-I. is grateful to A.~Yamada for
enlightening discussions.

\newpage

\appendix
\noindent
\begin{center}
{\Large {\bf Appendix}}
\end{center}

\section{Definitions}
\reseteqnum

On the $G_{224}$ stage,
the following representations of the fields
under $SU(2)_R\times SU(2)_L$
are employed:
\begin{eqnarray}
 \psi_R^T=(U_R\  D_R),\ \ \ \ \ \ \psi_L^T=(U_L\  D_L), \nonumber
\end{eqnarray}
\begin{eqnarray}
 \phi_{1, 15}=\left(
 \begin{array}{cc}
  \phi_1^0 & \phi_1^+\\
  \phi_2^- & \phi_2^0
 \end{array}
 \right)_{1,\ 15}, \hspace{2em}
 \phi_{10}=\phi_{10}^m \tau^m; \hspace{1em}
 \tau^m=\frac{1}{\sqrt{2}}\sigma^m, \hspace{1em}m=1, 2, 3
\end{eqnarray}
where $U$ and $D$ represent quartets of $SU(4)_C$ and $\sigma^m$ denote
Pauli matrices.
The definition of
$\widetilde{\phi}_{1, 15}$ is given by
\begin{eqnarray}
 \widetilde{\phi}_{1, 15} = \epsilon^T \phi_{1, 15}^*
 \epsilon;\ \ \ \ \
 \epsilon=\left(
 \begin{array}{cc}
  0 & 1 \\
  -1 & 0
 \end{array}
 \right).
\end{eqnarray}
The $SU(4)_{C}$ representations of
$\phi_{15}$ and $\phi_{10}$ are given by
\begin{eqnarray}
 \phi_{15}=\sum_{a=1}^{15}\phi_{15}^a T^a; \hspace{1em}
 {\rm tr}(T^aT^b)=\delta^{ab}, \hspace{1em}
 \sum_{a=1}^{15}T^aT^a=\frac{15}{4}{\bf 1}, \hspace{1em}
 \sum_{b=1}^{15}T^bT^aT^b=-\frac{1}{4}T^a, \nonumber
\end{eqnarray}
\begin{eqnarray}
 \phi_{10}=\sum_{\alpha=1}^{10}\phi_{10}^\alpha S_\alpha; \hspace{1em}
 {\rm tr}(S_\alpha S^\beta)=\delta_\alpha^\beta, \hspace{1em}
 \sum_{\alpha=1}^{10}S_\alpha S^\alpha=\frac{5}{2}{\bf 1},
 \hspace{1em} S^\alpha = S^*_\alpha,
\end{eqnarray}
where $T^a$ correspond to $SU(4)_C$ generators in the defining
representation.

\section{Quark Mass Matrices}
\reseteqnum

It seems instructive to
consider some analytical relations among
quark masses and mixings
which are expected to hold at the weak scale $M_W$ in our model.
We make a rough estimate that the running of each
Yukawa coupling between the scales $M_U$ and $M_W$
does not affect its order of magnitude,
which can be checked numerically.
Then the mass matrices at the weak scale can be written
in terms of rescaled couplings
\begin{eqnarray}
 y^u_{33} \sim y^d_{33} \sim \alpha y_{33}, \quad
 z^u_{22} \sim z^d_{22} \sim \frac{\gamma'}{2\sqrt{3}}z_{22},
 \quad {\it etc.}
\end{eqnarray}
as follows:
\begin{eqnarray}
 M_u &=&
  \left(
  \begin{array}{ccc}
  -\varepsilon^2 y_{11}^u       &
   \varepsilon^2 \ y_{12}^u        &
   \varepsilon (z^u_{13} + y_{13}^u)   \\
  \varepsilon^2 \xi y_{12}^u  &
   \varepsilon z_{22}^u   &
   -\varepsilon \xi y_{23}^u     \\
  \varepsilon (z_{13}^u - y_{13}^u)               &
   \varepsilon \xi y_{23}^u        &
   y_{33}^u
  \end{array}
  \right) v_W, \nonumber\\
 M_d &=&
  \left(
  \begin{array}{ccc}
  -\varepsilon^2 y_{11}^d       &
   \varepsilon^2 \xi^{-1} y_{12}^d        &
   \varepsilon (\xi^{-1} \zeta z^d_{13} + y_{13}^d)   \\
  \varepsilon^2 \xi^{-1} y_{12}^d  &
   \varepsilon \xi^{-1} \zeta z_{22}^d   &
   -\varepsilon \xi^{-1} y_{23}^d     \\
  \varepsilon (\xi^{-1} \zeta z_{13}^d - y_{13}^d)               &
   \varepsilon \xi^{-1} y_{23}^d        &
   y_{33}^d
  \end{array}
  \right) \xi v_W,
\end{eqnarray}
where
\begin{eqnarray}
 \xi = \frac{\beta}{\alpha}, \quad
 \zeta = \frac{\delta'}{\gamma'}.
\end{eqnarray}
Eq.(\ref{eq:rgesm})
implies that
the ratio between the Yukawa couplings
of top and bottom quarks
does not change considerably
\begin{eqnarray}
 y^u_{33} \simeq y^d_{33}
\end{eqnarray}
under dominance of the gauge couplings $g_2$ and $g_3$.

We now derive approximate relations for the parameters
$\xi$ and $\zeta$
with the aid of
smallness of the hierarchy factor
$\varepsilon = M_U/M_P \simeq 1/250$
(See section 3).
The masses of top and bottom quarks are
expressed as
\begin{eqnarray}
 \frac{m_t}{v_W} = y_{33}^u + {\rm O}(\varepsilon^2), \quad
 \frac{m_b}{\xi v_W} = y_{33}^d + {\rm O}(\varepsilon^2 \xi^{-2}),
\end{eqnarray}
which implies
\begin{eqnarray}
 \xi \simeq \frac{m_b}{m_t}.
\end{eqnarray}
If we put $m_t \simeq 160{\rm GeV}$,
we have
$\xi \simeq 1/50$,
which yields $\varepsilon \xi^{-1} \simeq 1/5$.

Let us proceed to the second generation.
We obtain
\begin{eqnarray}
 \frac{m_c}{v_W} = \varepsilon z_{22}^u
  + \varepsilon^2 \xi {{y^u_{23}}^2 \over y_{33}^u}
  + {\rm O}(\varepsilon^3), \quad
 \frac{m_s}{\xi v_W} = \varepsilon \xi^{-1} \zeta z_{22}^d
  + \varepsilon^2 \xi^{-2} {{y^d_{23}}^2 \over y_{33}^d}
  + {\rm O}(\varepsilon^3 \xi^{-3})
 \label{eq:secrel}
\end{eqnarray}
with a KM matrix element
\begin{eqnarray}
 V_{cb} = \varepsilon
  (\xi^{-1} \frac{y^d_{23}}{y^d_{33}}
   - \xi \frac{y^u_{23}}{y^u_{33}})
  + {\rm O}(\varepsilon^2 \xi^{-2}).
 \label{eq:cbelem}
\end{eqnarray}
Note that the O($\epsilon^2 \xi^{-2}$) term in eq.(\ref{eq:secrel})
is retained in anticipation of smallness of $\zeta$.
By means of these relations, we get
\begin{eqnarray}
 \zeta \simeq \frac{1}{m_c} (m_s - V_{cb}^2 m_b),
\end{eqnarray}
which is satisfied when $\zeta \simeq 1/10$.
The enhancement factor $\xi^{-1}$ for the mixing $V_{cb}$
in eq.(\ref{eq:cbelem})
comes from contribution of $\widetilde H$,
whose coupling is characteristic of non-supersymmetric models.
A choice $y_{23}^d/y_{33}^d \simeq 1/4$ yields
$V_{cb} \simeq 1/20$.


\newpage
\newcommand{\J}[4]{{\sl #1} {\bf #2} (19#3) #4}

\end{document}

%% file: gauge.tex
\setlength{\unitlength}{0.240900pt}
\ifx\plotpoint\undefined\newsavebox{\plotpoint}\fi
\sbox{\plotpoint}{\rule[-0.200pt]{0.400pt}{0.400pt}}%
\begin{picture}(1500,900)(0,0)
\font\gnuplot=cmr10 at 10pt
\gnuplot
\sbox{\plotpoint}{\rule[-0.200pt]{0.400pt}{0.400pt}}%
\put(220.0,113.0){\rule[-0.200pt]{292.934pt}{0.400pt}}
\put(220.0,113.0){\rule[-0.200pt]{0.400pt}{184.048pt}}
\put(220.0,113.0){\rule[-0.200pt]{292.934pt}{0.400pt}}
\put(220.0,113.0){\rule[-0.200pt]{4.818pt}{0.400pt}}
\put(198,113){\makebox(0,0)[r]{0}}
\put(1416.0,113.0){\rule[-0.200pt]{4.818pt}{0.400pt}}
\put(220.0,231.0){\rule[-0.200pt]{292.934pt}{0.400pt}}
\put(220.0,231.0){\rule[-0.200pt]{4.818pt}{0.400pt}}
\put(198,231){\makebox(0,0)[r]{10}}
\put(1416.0,231.0){\rule[-0.200pt]{4.818pt}{0.400pt}}
\put(220.0,348.0){\rule[-0.200pt]{292.934pt}{0.400pt}}
\put(220.0,348.0){\rule[-0.200pt]{4.818pt}{0.400pt}}
\put(198,348){\makebox(0,0)[r]{20}}
\put(1416.0,348.0){\rule[-0.200pt]{4.818pt}{0.400pt}}
\put(220.0,466.0){\rule[-0.200pt]{292.934pt}{0.400pt}}
\put(220.0,466.0){\rule[-0.200pt]{4.818pt}{0.400pt}}
\put(198,466){\makebox(0,0)[r]{30}}
\put(1416.0,466.0){\rule[-0.200pt]{4.818pt}{0.400pt}}
\put(220.0,583.0){\rule[-0.200pt]{292.934pt}{0.400pt}}
\put(220.0,583.0){\rule[-0.200pt]{4.818pt}{0.400pt}}
\put(198,583){\makebox(0,0)[r]{40}}
\put(1416.0,583.0){\rule[-0.200pt]{4.818pt}{0.400pt}}
\put(220.0,701.0){\rule[-0.200pt]{292.934pt}{0.400pt}}
\put(220.0,701.0){\rule[-0.200pt]{4.818pt}{0.400pt}}
\put(198,701){\makebox(0,0)[r]{50}}
\put(1416.0,701.0){\rule[-0.200pt]{4.818pt}{0.400pt}}
\put(220.0,818.0){\rule[-0.200pt]{292.934pt}{0.400pt}}
\put(220.0,818.0){\rule[-0.200pt]{4.818pt}{0.400pt}}
\put(198,818){\makebox(0,0)[r]{60}}
\put(1416.0,818.0){\rule[-0.200pt]{4.818pt}{0.400pt}}
\put(220.0,113.0){\rule[-0.200pt]{0.400pt}{184.048pt}}
\put(220.0,113.0){\rule[-0.200pt]{0.400pt}{4.818pt}}
\put(220,68){\makebox(0,0){0}}
\put(220.0,857.0){\rule[-0.200pt]{0.400pt}{4.818pt}}
\put(342.0,113.0){\rule[-0.200pt]{0.400pt}{184.048pt}}
\put(342.0,113.0){\rule[-0.200pt]{0.400pt}{4.818pt}}
\put(342,68){\makebox(0,0){2}}
\put(342.0,857.0){\rule[-0.200pt]{0.400pt}{4.818pt}}
\put(463.0,113.0){\rule[-0.200pt]{0.400pt}{184.048pt}}
\put(463.0,113.0){\rule[-0.200pt]{0.400pt}{4.818pt}}
\put(463,68){\makebox(0,0){4}}
\put(463.0,857.0){\rule[-0.200pt]{0.400pt}{4.818pt}}
\put(585.0,113.0){\rule[-0.200pt]{0.400pt}{184.048pt}}
\put(585.0,113.0){\rule[-0.200pt]{0.400pt}{4.818pt}}
\put(585,68){\makebox(0,0){6}}
\put(585.0,857.0){\rule[-0.200pt]{0.400pt}{4.818pt}}
\put(706.0,113.0){\rule[-0.200pt]{0.400pt}{184.048pt}}
\put(706.0,113.0){\rule[-0.200pt]{0.400pt}{4.818pt}}
\put(706,68){\makebox(0,0){8}}
\put(706.0,857.0){\rule[-0.200pt]{0.400pt}{4.818pt}}
\put(828.0,113.0){\rule[-0.200pt]{0.400pt}{184.048pt}}
\put(828.0,113.0){\rule[-0.200pt]{0.400pt}{4.818pt}}
\put(828,68){\makebox(0,0){10}}
\put(828.0,857.0){\rule[-0.200pt]{0.400pt}{4.818pt}}
\put(950.0,113.0){\rule[-0.200pt]{0.400pt}{184.048pt}}
\put(950.0,113.0){\rule[-0.200pt]{0.400pt}{4.818pt}}
\put(950,68){\makebox(0,0){12}}
\put(950.0,857.0){\rule[-0.200pt]{0.400pt}{4.818pt}}
\put(1071.0,113.0){\rule[-0.200pt]{0.400pt}{184.048pt}}
\put(1071.0,113.0){\rule[-0.200pt]{0.400pt}{4.818pt}}
\put(1071,68){\makebox(0,0){14}}
\put(1071.0,857.0){\rule[-0.200pt]{0.400pt}{4.818pt}}
\put(1193.0,113.0){\rule[-0.200pt]{0.400pt}{184.048pt}}
\put(1193.0,113.0){\rule[-0.200pt]{0.400pt}{4.818pt}}
\put(1193,68){\makebox(0,0){16}}
\put(1193.0,857.0){\rule[-0.200pt]{0.400pt}{4.818pt}}
\put(1314.0,113.0){\rule[-0.200pt]{0.400pt}{184.048pt}}
\put(1314.0,113.0){\rule[-0.200pt]{0.400pt}{4.818pt}}
\put(1314,68){\makebox(0,0){18}}
\put(1314.0,857.0){\rule[-0.200pt]{0.400pt}{4.818pt}}
\put(1436.0,113.0){\rule[-0.200pt]{0.400pt}{184.048pt}}
\put(1436.0,113.0){\rule[-0.200pt]{0.400pt}{4.818pt}}
\put(1436,68){\makebox(0,0){20}}
\put(1436.0,857.0){\rule[-0.200pt]{0.400pt}{4.818pt}}
\put(220.0,113.0){\rule[-0.200pt]{292.934pt}{0.400pt}}
\put(1436.0,113.0){\rule[-0.200pt]{0.400pt}{184.048pt}}
\put(220.0,877.0){\rule[-0.200pt]{292.934pt}{0.400pt}}
\put(45,495){\makebox(0,0){$\omega$}}
\put(828,23){\makebox(0,0){$t={\rm log}_{10}\mu$}}
\put(524,341){\makebox(0,0){$\omega_{3C}$}}
\put(524,547){\makebox(0,0){$\omega_{L}$}}
\put(524,794){\makebox(0,0){$\omega_{Y}$}}
\put(998,536){\makebox(0,0){$\omega_{4C}$}}
\put(998,620){\makebox(0,0){$\omega_{L}$}}
\put(998,748){\makebox(0,0){$\omega_{R}$}}
\put(1375,542){\makebox(0,0){$\omega_{U}$}}
\put(220.0,113.0){\rule[-0.200pt]{0.400pt}{184.048pt}}
\put(220,844){\usebox{\plotpoint}}
\multiput(220.00,842.93)(1.814,-0.488){13}{\rule{1.500pt}{0.117pt}}
\multiput(220.00,843.17)(24.887,-8.000){2}{\rule{0.750pt}{0.400pt}}
\multiput(248.00,834.93)(1.601,-0.489){15}{\rule{1.344pt}{0.118pt}}
\multiput(248.00,835.17)(25.210,-9.000){2}{\rule{0.672pt}{0.400pt}}
\multiput(276.00,825.93)(1.814,-0.488){13}{\rule{1.500pt}{0.117pt}}
\multiput(276.00,826.17)(24.887,-8.000){2}{\rule{0.750pt}{0.400pt}}
\multiput(304.00,817.93)(1.814,-0.488){13}{\rule{1.500pt}{0.117pt}}
\multiput(304.00,818.17)(24.887,-8.000){2}{\rule{0.750pt}{0.400pt}}
\multiput(332.00,809.93)(1.814,-0.488){13}{\rule{1.500pt}{0.117pt}}
\multiput(332.00,810.17)(24.887,-8.000){2}{\rule{0.750pt}{0.400pt}}
\multiput(360.00,801.93)(1.814,-0.488){13}{\rule{1.500pt}{0.117pt}}
\multiput(360.00,802.17)(24.887,-8.000){2}{\rule{0.750pt}{0.400pt}}
\multiput(388.00,793.93)(1.814,-0.488){13}{\rule{1.500pt}{0.117pt}}
\multiput(388.00,794.17)(24.887,-8.000){2}{\rule{0.750pt}{0.400pt}}
\multiput(416.00,785.93)(1.814,-0.488){13}{\rule{1.500pt}{0.117pt}}
\multiput(416.00,786.17)(24.887,-8.000){2}{\rule{0.750pt}{0.400pt}}
\multiput(444.00,777.93)(1.748,-0.488){13}{\rule{1.450pt}{0.117pt}}
\multiput(444.00,778.17)(23.990,-8.000){2}{\rule{0.725pt}{0.400pt}}
\multiput(471.00,769.93)(1.601,-0.489){15}{\rule{1.344pt}{0.118pt}}
\multiput(471.00,770.17)(25.210,-9.000){2}{\rule{0.672pt}{0.400pt}}
\multiput(499.00,760.93)(1.814,-0.488){13}{\rule{1.500pt}{0.117pt}}
\multiput(499.00,761.17)(24.887,-8.000){2}{\rule{0.750pt}{0.400pt}}
\multiput(527.00,752.93)(1.814,-0.488){13}{\rule{1.500pt}{0.117pt}}
\multiput(527.00,753.17)(24.887,-8.000){2}{\rule{0.750pt}{0.400pt}}
\multiput(555.00,744.93)(1.814,-0.488){13}{\rule{1.500pt}{0.117pt}}
\multiput(555.00,745.17)(24.887,-8.000){2}{\rule{0.750pt}{0.400pt}}
\multiput(583.00,736.93)(1.814,-0.488){13}{\rule{1.500pt}{0.117pt}}
\multiput(583.00,737.17)(24.887,-8.000){2}{\rule{0.750pt}{0.400pt}}
\multiput(611.00,728.93)(1.814,-0.488){13}{\rule{1.500pt}{0.117pt}}
\multiput(611.00,729.17)(24.887,-8.000){2}{\rule{0.750pt}{0.400pt}}
\multiput(639.00,720.93)(1.814,-0.488){13}{\rule{1.500pt}{0.117pt}}
\multiput(639.00,721.17)(24.887,-8.000){2}{\rule{0.750pt}{0.400pt}}
\multiput(667.00,712.93)(1.814,-0.488){13}{\rule{1.500pt}{0.117pt}}
\multiput(667.00,713.17)(24.887,-8.000){2}{\rule{0.750pt}{0.400pt}}
\multiput(695.00,704.93)(1.814,-0.488){13}{\rule{1.500pt}{0.117pt}}
\multiput(695.00,705.17)(24.887,-8.000){2}{\rule{0.750pt}{0.400pt}}
\multiput(723.00,696.93)(1.601,-0.489){15}{\rule{1.344pt}{0.118pt}}
\multiput(723.00,697.17)(25.210,-9.000){2}{\rule{0.672pt}{0.400pt}}
\multiput(751.00,687.93)(1.814,-0.488){13}{\rule{1.500pt}{0.117pt}}
\multiput(751.00,688.17)(24.887,-8.000){2}{\rule{0.750pt}{0.400pt}}
\multiput(779.00,679.93)(1.814,-0.488){13}{\rule{1.500pt}{0.117pt}}
\multiput(779.00,680.17)(24.887,-8.000){2}{\rule{0.750pt}{0.400pt}}
\multiput(807.00,671.93)(1.814,-0.488){13}{\rule{1.500pt}{0.117pt}}
\multiput(807.00,672.17)(24.887,-8.000){2}{\rule{0.750pt}{0.400pt}}
\multiput(835.00,663.93)(1.814,-0.488){13}{\rule{1.500pt}{0.117pt}}
\multiput(835.00,664.17)(24.887,-8.000){2}{\rule{0.750pt}{0.400pt}}
\multiput(863.00,655.93)(1.814,-0.488){13}{\rule{1.500pt}{0.117pt}}
\multiput(863.00,656.17)(24.887,-8.000){2}{\rule{0.750pt}{0.400pt}}
\put(220,440){\usebox{\plotpoint}}
\multiput(220.00,440.59)(2.480,0.482){9}{\rule{1.967pt}{0.116pt}}
\multiput(220.00,439.17)(23.918,6.000){2}{\rule{0.983pt}{0.400pt}}
\multiput(248.00,446.59)(2.094,0.485){11}{\rule{1.700pt}{0.117pt}}
\multiput(248.00,445.17)(24.472,7.000){2}{\rule{0.850pt}{0.400pt}}
\multiput(276.00,453.59)(2.480,0.482){9}{\rule{1.967pt}{0.116pt}}
\multiput(276.00,452.17)(23.918,6.000){2}{\rule{0.983pt}{0.400pt}}
\multiput(304.00,459.59)(2.480,0.482){9}{\rule{1.967pt}{0.116pt}}
\multiput(304.00,458.17)(23.918,6.000){2}{\rule{0.983pt}{0.400pt}}
\multiput(332.00,465.59)(2.480,0.482){9}{\rule{1.967pt}{0.116pt}}
\multiput(332.00,464.17)(23.918,6.000){2}{\rule{0.983pt}{0.400pt}}
\multiput(360.00,471.59)(2.094,0.485){11}{\rule{1.700pt}{0.117pt}}
\multiput(360.00,470.17)(24.472,7.000){2}{\rule{0.850pt}{0.400pt}}
\multiput(388.00,478.59)(2.480,0.482){9}{\rule{1.967pt}{0.116pt}}
\multiput(388.00,477.17)(23.918,6.000){2}{\rule{0.983pt}{0.400pt}}
\multiput(416.00,484.59)(2.480,0.482){9}{\rule{1.967pt}{0.116pt}}
\multiput(416.00,483.17)(23.918,6.000){2}{\rule{0.983pt}{0.400pt}}
\multiput(444.00,490.59)(2.389,0.482){9}{\rule{1.900pt}{0.116pt}}
\multiput(444.00,489.17)(23.056,6.000){2}{\rule{0.950pt}{0.400pt}}
\multiput(471.00,496.59)(2.094,0.485){11}{\rule{1.700pt}{0.117pt}}
\multiput(471.00,495.17)(24.472,7.000){2}{\rule{0.850pt}{0.400pt}}
\multiput(499.00,503.59)(2.480,0.482){9}{\rule{1.967pt}{0.116pt}}
\multiput(499.00,502.17)(23.918,6.000){2}{\rule{0.983pt}{0.400pt}}
\multiput(527.00,509.59)(2.480,0.482){9}{\rule{1.967pt}{0.116pt}}
\multiput(527.00,508.17)(23.918,6.000){2}{\rule{0.983pt}{0.400pt}}
\multiput(555.00,515.59)(2.094,0.485){11}{\rule{1.700pt}{0.117pt}}
\multiput(555.00,514.17)(24.472,7.000){2}{\rule{0.850pt}{0.400pt}}
\multiput(583.00,522.59)(2.480,0.482){9}{\rule{1.967pt}{0.116pt}}
\multiput(583.00,521.17)(23.918,6.000){2}{\rule{0.983pt}{0.400pt}}
\multiput(611.00,528.59)(2.480,0.482){9}{\rule{1.967pt}{0.116pt}}
\multiput(611.00,527.17)(23.918,6.000){2}{\rule{0.983pt}{0.400pt}}
\multiput(639.00,534.59)(2.480,0.482){9}{\rule{1.967pt}{0.116pt}}
\multiput(639.00,533.17)(23.918,6.000){2}{\rule{0.983pt}{0.400pt}}
\multiput(667.00,540.59)(2.094,0.485){11}{\rule{1.700pt}{0.117pt}}
\multiput(667.00,539.17)(24.472,7.000){2}{\rule{0.850pt}{0.400pt}}
\multiput(695.00,547.59)(2.480,0.482){9}{\rule{1.967pt}{0.116pt}}
\multiput(695.00,546.17)(23.918,6.000){2}{\rule{0.983pt}{0.400pt}}
\multiput(723.00,553.59)(2.480,0.482){9}{\rule{1.967pt}{0.116pt}}
\multiput(723.00,552.17)(23.918,6.000){2}{\rule{0.983pt}{0.400pt}}
\multiput(751.00,559.59)(2.480,0.482){9}{\rule{1.967pt}{0.116pt}}
\multiput(751.00,558.17)(23.918,6.000){2}{\rule{0.983pt}{0.400pt}}
\multiput(779.00,565.59)(2.094,0.485){11}{\rule{1.700pt}{0.117pt}}
\multiput(779.00,564.17)(24.472,7.000){2}{\rule{0.850pt}{0.400pt}}
\multiput(807.00,572.59)(2.480,0.482){9}{\rule{1.967pt}{0.116pt}}
\multiput(807.00,571.17)(23.918,6.000){2}{\rule{0.983pt}{0.400pt}}
\multiput(835.00,578.59)(2.480,0.482){9}{\rule{1.967pt}{0.116pt}}
\multiput(835.00,577.17)(23.918,6.000){2}{\rule{0.983pt}{0.400pt}}
\multiput(863.00,584.59)(2.094,0.485){11}{\rule{1.700pt}{0.117pt}}
\multiput(863.00,583.17)(24.472,7.000){2}{\rule{0.850pt}{0.400pt}}
\put(220,146){\usebox{\plotpoint}}
\multiput(220.00,146.58)(1.011,0.494){25}{\rule{0.900pt}{0.119pt}}
\multiput(220.00,145.17)(26.132,14.000){2}{\rule{0.450pt}{0.400pt}}
\multiput(248.00,160.58)(1.011,0.494){25}{\rule{0.900pt}{0.119pt}}
\multiput(248.00,159.17)(26.132,14.000){2}{\rule{0.450pt}{0.400pt}}
\multiput(276.00,174.58)(1.011,0.494){25}{\rule{0.900pt}{0.119pt}}
\multiput(276.00,173.17)(26.132,14.000){2}{\rule{0.450pt}{0.400pt}}
\multiput(304.00,188.58)(1.011,0.494){25}{\rule{0.900pt}{0.119pt}}
\multiput(304.00,187.17)(26.132,14.000){2}{\rule{0.450pt}{0.400pt}}
\multiput(332.00,202.58)(1.011,0.494){25}{\rule{0.900pt}{0.119pt}}
\multiput(332.00,201.17)(26.132,14.000){2}{\rule{0.450pt}{0.400pt}}
\multiput(360.00,216.58)(1.011,0.494){25}{\rule{0.900pt}{0.119pt}}
\multiput(360.00,215.17)(26.132,14.000){2}{\rule{0.450pt}{0.400pt}}
\multiput(388.00,230.58)(1.091,0.493){23}{\rule{0.962pt}{0.119pt}}
\multiput(388.00,229.17)(26.004,13.000){2}{\rule{0.481pt}{0.400pt}}
\multiput(416.00,243.58)(1.011,0.494){25}{\rule{0.900pt}{0.119pt}}
\multiput(416.00,242.17)(26.132,14.000){2}{\rule{0.450pt}{0.400pt}}
\multiput(444.00,257.58)(0.974,0.494){25}{\rule{0.871pt}{0.119pt}}
\multiput(444.00,256.17)(25.191,14.000){2}{\rule{0.436pt}{0.400pt}}
\multiput(471.00,271.58)(1.011,0.494){25}{\rule{0.900pt}{0.119pt}}
\multiput(471.00,270.17)(26.132,14.000){2}{\rule{0.450pt}{0.400pt}}
\multiput(499.00,285.58)(1.011,0.494){25}{\rule{0.900pt}{0.119pt}}
\multiput(499.00,284.17)(26.132,14.000){2}{\rule{0.450pt}{0.400pt}}
\multiput(527.00,299.58)(1.011,0.494){25}{\rule{0.900pt}{0.119pt}}
\multiput(527.00,298.17)(26.132,14.000){2}{\rule{0.450pt}{0.400pt}}
\multiput(555.00,313.58)(1.011,0.494){25}{\rule{0.900pt}{0.119pt}}
\multiput(555.00,312.17)(26.132,14.000){2}{\rule{0.450pt}{0.400pt}}
\multiput(583.00,327.58)(1.091,0.493){23}{\rule{0.962pt}{0.119pt}}
\multiput(583.00,326.17)(26.004,13.000){2}{\rule{0.481pt}{0.400pt}}
\multiput(611.00,340.58)(1.011,0.494){25}{\rule{0.900pt}{0.119pt}}
\multiput(611.00,339.17)(26.132,14.000){2}{\rule{0.450pt}{0.400pt}}
\multiput(639.00,354.58)(1.011,0.494){25}{\rule{0.900pt}{0.119pt}}
\multiput(639.00,353.17)(26.132,14.000){2}{\rule{0.450pt}{0.400pt}}
\multiput(667.00,368.58)(1.011,0.494){25}{\rule{0.900pt}{0.119pt}}
\multiput(667.00,367.17)(26.132,14.000){2}{\rule{0.450pt}{0.400pt}}
\multiput(695.00,382.58)(1.011,0.494){25}{\rule{0.900pt}{0.119pt}}
\multiput(695.00,381.17)(26.132,14.000){2}{\rule{0.450pt}{0.400pt}}
\multiput(723.00,396.58)(1.011,0.494){25}{\rule{0.900pt}{0.119pt}}
\multiput(723.00,395.17)(26.132,14.000){2}{\rule{0.450pt}{0.400pt}}
\multiput(751.00,410.58)(1.011,0.494){25}{\rule{0.900pt}{0.119pt}}
\multiput(751.00,409.17)(26.132,14.000){2}{\rule{0.450pt}{0.400pt}}
\multiput(779.00,424.58)(1.091,0.493){23}{\rule{0.962pt}{0.119pt}}
\multiput(779.00,423.17)(26.004,13.000){2}{\rule{0.481pt}{0.400pt}}
\multiput(807.00,437.58)(1.011,0.494){25}{\rule{0.900pt}{0.119pt}}
\multiput(807.00,436.17)(26.132,14.000){2}{\rule{0.450pt}{0.400pt}}
\multiput(835.00,451.58)(1.011,0.494){25}{\rule{0.900pt}{0.119pt}}
\multiput(835.00,450.17)(26.132,14.000){2}{\rule{0.450pt}{0.400pt}}
\multiput(863.00,465.58)(1.011,0.494){25}{\rule{0.900pt}{0.119pt}}
\multiput(863.00,464.17)(26.132,14.000){2}{\rule{0.450pt}{0.400pt}}
\put(891,591){\usebox{\plotpoint}}
\multiput(891.00,589.95)(3.141,-0.447){3}{\rule{2.100pt}{0.108pt}}
\multiput(891.00,590.17)(10.641,-3.000){2}{\rule{1.050pt}{0.400pt}}
\put(906,586.17){\rule{3.100pt}{0.400pt}}
\multiput(906.00,587.17)(8.566,-2.000){2}{\rule{1.550pt}{0.400pt}}
\put(921,584.17){\rule{3.100pt}{0.400pt}}
\multiput(921.00,585.17)(8.566,-2.000){2}{\rule{1.550pt}{0.400pt}}
\put(936,582.17){\rule{3.100pt}{0.400pt}}
\multiput(936.00,583.17)(8.566,-2.000){2}{\rule{1.550pt}{0.400pt}}
\put(951,580.17){\rule{3.100pt}{0.400pt}}
\multiput(951.00,581.17)(8.566,-2.000){2}{\rule{1.550pt}{0.400pt}}
\put(966,578.17){\rule{3.100pt}{0.400pt}}
\multiput(966.00,579.17)(8.566,-2.000){2}{\rule{1.550pt}{0.400pt}}
\put(981,576.17){\rule{3.100pt}{0.400pt}}
\multiput(981.00,577.17)(8.566,-2.000){2}{\rule{1.550pt}{0.400pt}}
\put(996,574.17){\rule{3.100pt}{0.400pt}}
\multiput(996.00,575.17)(8.566,-2.000){2}{\rule{1.550pt}{0.400pt}}
\multiput(1011.00,572.95)(3.141,-0.447){3}{\rule{2.100pt}{0.108pt}}
\multiput(1011.00,573.17)(10.641,-3.000){2}{\rule{1.050pt}{0.400pt}}
\put(1026,569.17){\rule{3.100pt}{0.400pt}}
\multiput(1026.00,570.17)(8.566,-2.000){2}{\rule{1.550pt}{0.400pt}}
\put(1041,567.17){\rule{3.100pt}{0.400pt}}
\multiput(1041.00,568.17)(8.566,-2.000){2}{\rule{1.550pt}{0.400pt}}
\put(1056,565.17){\rule{3.100pt}{0.400pt}}
\multiput(1056.00,566.17)(8.566,-2.000){2}{\rule{1.550pt}{0.400pt}}
\put(1071,563.17){\rule{3.100pt}{0.400pt}}
\multiput(1071.00,564.17)(8.566,-2.000){2}{\rule{1.550pt}{0.400pt}}
\put(1086,561.17){\rule{3.100pt}{0.400pt}}
\multiput(1086.00,562.17)(8.566,-2.000){2}{\rule{1.550pt}{0.400pt}}
\put(1101,559.17){\rule{3.100pt}{0.400pt}}
\multiput(1101.00,560.17)(8.566,-2.000){2}{\rule{1.550pt}{0.400pt}}
\put(1116,557.17){\rule{3.100pt}{0.400pt}}
\multiput(1116.00,558.17)(8.566,-2.000){2}{\rule{1.550pt}{0.400pt}}
\multiput(1131.00,555.95)(3.141,-0.447){3}{\rule{2.100pt}{0.108pt}}
\multiput(1131.00,556.17)(10.641,-3.000){2}{\rule{1.050pt}{0.400pt}}
\put(1146,552.17){\rule{3.100pt}{0.400pt}}
\multiput(1146.00,553.17)(8.566,-2.000){2}{\rule{1.550pt}{0.400pt}}
\put(1161,550.17){\rule{3.100pt}{0.400pt}}
\multiput(1161.00,551.17)(8.566,-2.000){2}{\rule{1.550pt}{0.400pt}}
\put(1176,548.17){\rule{3.100pt}{0.400pt}}
\multiput(1176.00,549.17)(8.566,-2.000){2}{\rule{1.550pt}{0.400pt}}
\put(1191,546.17){\rule{3.100pt}{0.400pt}}
\multiput(1191.00,547.17)(8.566,-2.000){2}{\rule{1.550pt}{0.400pt}}
\put(1206,544.17){\rule{3.100pt}{0.400pt}}
\multiput(1206.00,545.17)(8.566,-2.000){2}{\rule{1.550pt}{0.400pt}}
\put(1221,542.17){\rule{3.100pt}{0.400pt}}
\multiput(1221.00,543.17)(8.566,-2.000){2}{\rule{1.550pt}{0.400pt}}
\put(1236,540.17){\rule{3.100pt}{0.400pt}}
\multiput(1236.00,541.17)(8.566,-2.000){2}{\rule{1.550pt}{0.400pt}}
\put(891,762){\usebox{\plotpoint}}
\multiput(891.00,760.93)(0.844,-0.489){15}{\rule{0.767pt}{0.118pt}}
\multiput(891.00,761.17)(13.409,-9.000){2}{\rule{0.383pt}{0.400pt}}
\multiput(906.00,751.93)(0.844,-0.489){15}{\rule{0.767pt}{0.118pt}}
\multiput(906.00,752.17)(13.409,-9.000){2}{\rule{0.383pt}{0.400pt}}
\multiput(921.00,742.92)(0.756,-0.491){17}{\rule{0.700pt}{0.118pt}}
\multiput(921.00,743.17)(13.547,-10.000){2}{\rule{0.350pt}{0.400pt}}
\multiput(936.00,732.93)(0.844,-0.489){15}{\rule{0.767pt}{0.118pt}}
\multiput(936.00,733.17)(13.409,-9.000){2}{\rule{0.383pt}{0.400pt}}
\multiput(951.00,723.93)(0.844,-0.489){15}{\rule{0.767pt}{0.118pt}}
\multiput(951.00,724.17)(13.409,-9.000){2}{\rule{0.383pt}{0.400pt}}
\multiput(966.00,714.93)(0.844,-0.489){15}{\rule{0.767pt}{0.118pt}}
\multiput(966.00,715.17)(13.409,-9.000){2}{\rule{0.383pt}{0.400pt}}
\multiput(981.00,705.93)(0.844,-0.489){15}{\rule{0.767pt}{0.118pt}}
\multiput(981.00,706.17)(13.409,-9.000){2}{\rule{0.383pt}{0.400pt}}
\multiput(996.00,696.92)(0.756,-0.491){17}{\rule{0.700pt}{0.118pt}}
\multiput(996.00,697.17)(13.547,-10.000){2}{\rule{0.350pt}{0.400pt}}
\multiput(1011.00,686.93)(0.844,-0.489){15}{\rule{0.767pt}{0.118pt}}
\multiput(1011.00,687.17)(13.409,-9.000){2}{\rule{0.383pt}{0.400pt}}
\multiput(1026.00,677.93)(0.844,-0.489){15}{\rule{0.767pt}{0.118pt}}
\multiput(1026.00,678.17)(13.409,-9.000){2}{\rule{0.383pt}{0.400pt}}
\multiput(1041.00,668.93)(0.844,-0.489){15}{\rule{0.767pt}{0.118pt}}
\multiput(1041.00,669.17)(13.409,-9.000){2}{\rule{0.383pt}{0.400pt}}
\multiput(1056.00,659.93)(0.844,-0.489){15}{\rule{0.767pt}{0.118pt}}
\multiput(1056.00,660.17)(13.409,-9.000){2}{\rule{0.383pt}{0.400pt}}
\multiput(1071.00,650.92)(0.756,-0.491){17}{\rule{0.700pt}{0.118pt}}
\multiput(1071.00,651.17)(13.547,-10.000){2}{\rule{0.350pt}{0.400pt}}
\multiput(1086.00,640.93)(0.844,-0.489){15}{\rule{0.767pt}{0.118pt}}
\multiput(1086.00,641.17)(13.409,-9.000){2}{\rule{0.383pt}{0.400pt}}
\multiput(1101.00,631.93)(0.844,-0.489){15}{\rule{0.767pt}{0.118pt}}
\multiput(1101.00,632.17)(13.409,-9.000){2}{\rule{0.383pt}{0.400pt}}
\multiput(1116.00,622.93)(0.844,-0.489){15}{\rule{0.767pt}{0.118pt}}
\multiput(1116.00,623.17)(13.409,-9.000){2}{\rule{0.383pt}{0.400pt}}
\multiput(1131.00,613.93)(0.844,-0.489){15}{\rule{0.767pt}{0.118pt}}
\multiput(1131.00,614.17)(13.409,-9.000){2}{\rule{0.383pt}{0.400pt}}
\multiput(1146.00,604.92)(0.756,-0.491){17}{\rule{0.700pt}{0.118pt}}
\multiput(1146.00,605.17)(13.547,-10.000){2}{\rule{0.350pt}{0.400pt}}
\multiput(1161.00,594.93)(0.844,-0.489){15}{\rule{0.767pt}{0.118pt}}
\multiput(1161.00,595.17)(13.409,-9.000){2}{\rule{0.383pt}{0.400pt}}
\multiput(1176.00,585.93)(0.844,-0.489){15}{\rule{0.767pt}{0.118pt}}
\multiput(1176.00,586.17)(13.409,-9.000){2}{\rule{0.383pt}{0.400pt}}
\multiput(1191.00,576.93)(0.844,-0.489){15}{\rule{0.767pt}{0.118pt}}
\multiput(1191.00,577.17)(13.409,-9.000){2}{\rule{0.383pt}{0.400pt}}
\multiput(1206.00,567.92)(0.756,-0.491){17}{\rule{0.700pt}{0.118pt}}
\multiput(1206.00,568.17)(13.547,-10.000){2}{\rule{0.350pt}{0.400pt}}
\multiput(1221.00,557.93)(0.844,-0.489){15}{\rule{0.767pt}{0.118pt}}
\multiput(1221.00,558.17)(13.409,-9.000){2}{\rule{0.383pt}{0.400pt}}
\multiput(1236.00,548.93)(0.844,-0.489){15}{\rule{0.767pt}{0.118pt}}
\multiput(1236.00,549.17)(13.409,-9.000){2}{\rule{0.383pt}{0.400pt}}
\put(891,479){\usebox{\plotpoint}}
\put(891,479.17){\rule{3.100pt}{0.400pt}}
\multiput(891.00,478.17)(8.566,2.000){2}{\rule{1.550pt}{0.400pt}}
\multiput(906.00,481.61)(3.141,0.447){3}{\rule{2.100pt}{0.108pt}}
\multiput(906.00,480.17)(10.641,3.000){2}{\rule{1.050pt}{0.400pt}}
\put(921,484.17){\rule{3.100pt}{0.400pt}}
\multiput(921.00,483.17)(8.566,2.000){2}{\rule{1.550pt}{0.400pt}}
\multiput(936.00,486.61)(3.141,0.447){3}{\rule{2.100pt}{0.108pt}}
\multiput(936.00,485.17)(10.641,3.000){2}{\rule{1.050pt}{0.400pt}}
\put(951,489.17){\rule{3.100pt}{0.400pt}}
\multiput(951.00,488.17)(8.566,2.000){2}{\rule{1.550pt}{0.400pt}}
\multiput(966.00,491.61)(3.141,0.447){3}{\rule{2.100pt}{0.108pt}}
\multiput(966.00,490.17)(10.641,3.000){2}{\rule{1.050pt}{0.400pt}}
\put(981,494.17){\rule{3.100pt}{0.400pt}}
\multiput(981.00,493.17)(8.566,2.000){2}{\rule{1.550pt}{0.400pt}}
\multiput(996.00,496.61)(3.141,0.447){3}{\rule{2.100pt}{0.108pt}}
\multiput(996.00,495.17)(10.641,3.000){2}{\rule{1.050pt}{0.400pt}}
\put(1011,499.17){\rule{3.100pt}{0.400pt}}
\multiput(1011.00,498.17)(8.566,2.000){2}{\rule{1.550pt}{0.400pt}}
\multiput(1026.00,501.61)(3.141,0.447){3}{\rule{2.100pt}{0.108pt}}
\multiput(1026.00,500.17)(10.641,3.000){2}{\rule{1.050pt}{0.400pt}}
\put(1041,504.17){\rule{3.100pt}{0.400pt}}
\multiput(1041.00,503.17)(8.566,2.000){2}{\rule{1.550pt}{0.400pt}}
\multiput(1056.00,506.61)(3.141,0.447){3}{\rule{2.100pt}{0.108pt}}
\multiput(1056.00,505.17)(10.641,3.000){2}{\rule{1.050pt}{0.400pt}}
\put(1071,509.17){\rule{3.100pt}{0.400pt}}
\multiput(1071.00,508.17)(8.566,2.000){2}{\rule{1.550pt}{0.400pt}}
\multiput(1086.00,511.61)(3.141,0.447){3}{\rule{2.100pt}{0.108pt}}
\multiput(1086.00,510.17)(10.641,3.000){2}{\rule{1.050pt}{0.400pt}}
\put(1101,514.17){\rule{3.100pt}{0.400pt}}
\multiput(1101.00,513.17)(8.566,2.000){2}{\rule{1.550pt}{0.400pt}}
\multiput(1116.00,516.61)(3.141,0.447){3}{\rule{2.100pt}{0.108pt}}
\multiput(1116.00,515.17)(10.641,3.000){2}{\rule{1.050pt}{0.400pt}}
\put(1131,519.17){\rule{3.100pt}{0.400pt}}
\multiput(1131.00,518.17)(8.566,2.000){2}{\rule{1.550pt}{0.400pt}}
\multiput(1146.00,521.61)(3.141,0.447){3}{\rule{2.100pt}{0.108pt}}
\multiput(1146.00,520.17)(10.641,3.000){2}{\rule{1.050pt}{0.400pt}}
\put(1161,524.17){\rule{3.100pt}{0.400pt}}
\multiput(1161.00,523.17)(8.566,2.000){2}{\rule{1.550pt}{0.400pt}}
\multiput(1176.00,526.61)(3.141,0.447){3}{\rule{2.100pt}{0.108pt}}
\multiput(1176.00,525.17)(10.641,3.000){2}{\rule{1.050pt}{0.400pt}}
\put(1191,529.17){\rule{3.100pt}{0.400pt}}
\multiput(1191.00,528.17)(8.566,2.000){2}{\rule{1.550pt}{0.400pt}}
\multiput(1206.00,531.61)(3.141,0.447){3}{\rule{2.100pt}{0.108pt}}
\multiput(1206.00,530.17)(10.641,3.000){2}{\rule{1.050pt}{0.400pt}}
\put(1221,534.17){\rule{3.100pt}{0.400pt}}
\multiput(1221.00,533.17)(8.566,2.000){2}{\rule{1.550pt}{0.400pt}}
\multiput(1236.00,536.61)(3.141,0.447){3}{\rule{2.100pt}{0.108pt}}
\multiput(1236.00,535.17)(10.641,3.000){2}{\rule{1.050pt}{0.400pt}}
\put(1251,541){\usebox{\plotpoint}}
\multiput(1251.00,539.95)(1.355,-0.447){3}{\rule{1.033pt}{0.108pt}}
\multiput(1251.00,540.17)(4.855,-3.000){2}{\rule{0.517pt}{0.400pt}}
\multiput(1258.00,536.95)(1.579,-0.447){3}{\rule{1.167pt}{0.108pt}}
\multiput(1258.00,537.17)(5.579,-3.000){2}{\rule{0.583pt}{0.400pt}}
\multiput(1266.00,533.95)(1.579,-0.447){3}{\rule{1.167pt}{0.108pt}}
\multiput(1266.00,534.17)(5.579,-3.000){2}{\rule{0.583pt}{0.400pt}}
\multiput(1274.00,530.95)(1.355,-0.447){3}{\rule{1.033pt}{0.108pt}}
\multiput(1274.00,531.17)(4.855,-3.000){2}{\rule{0.517pt}{0.400pt}}
\multiput(1281.00,527.95)(1.579,-0.447){3}{\rule{1.167pt}{0.108pt}}
\multiput(1281.00,528.17)(5.579,-3.000){2}{\rule{0.583pt}{0.400pt}}
\put(1289,524.17){\rule{1.700pt}{0.400pt}}
\multiput(1289.00,525.17)(4.472,-2.000){2}{\rule{0.850pt}{0.400pt}}
\multiput(1297.00,522.95)(1.579,-0.447){3}{\rule{1.167pt}{0.108pt}}
\multiput(1297.00,523.17)(5.579,-3.000){2}{\rule{0.583pt}{0.400pt}}
\multiput(1305.00,519.95)(1.355,-0.447){3}{\rule{1.033pt}{0.108pt}}
\multiput(1305.00,520.17)(4.855,-3.000){2}{\rule{0.517pt}{0.400pt}}
\multiput(1312.00,516.95)(1.579,-0.447){3}{\rule{1.167pt}{0.108pt}}
\multiput(1312.00,517.17)(5.579,-3.000){2}{\rule{0.583pt}{0.400pt}}
\multiput(1320.00,513.95)(1.579,-0.447){3}{\rule{1.167pt}{0.108pt}}
\multiput(1320.00,514.17)(5.579,-3.000){2}{\rule{0.583pt}{0.400pt}}
\multiput(1328.00,510.95)(1.579,-0.447){3}{\rule{1.167pt}{0.108pt}}
\multiput(1328.00,511.17)(5.579,-3.000){2}{\rule{0.583pt}{0.400pt}}
\multiput(1336.00,507.95)(1.355,-0.447){3}{\rule{1.033pt}{0.108pt}}
\multiput(1336.00,508.17)(4.855,-3.000){2}{\rule{0.517pt}{0.400pt}}
\multiput(1343.00,504.95)(1.579,-0.447){3}{\rule{1.167pt}{0.108pt}}
\multiput(1343.00,505.17)(5.579,-3.000){2}{\rule{0.583pt}{0.400pt}}
\multiput(1351.00,501.95)(1.579,-0.447){3}{\rule{1.167pt}{0.108pt}}
\multiput(1351.00,502.17)(5.579,-3.000){2}{\rule{0.583pt}{0.400pt}}
\multiput(1359.00,498.95)(1.355,-0.447){3}{\rule{1.033pt}{0.108pt}}
\multiput(1359.00,499.17)(4.855,-3.000){2}{\rule{0.517pt}{0.400pt}}
\multiput(1366.00,495.95)(1.579,-0.447){3}{\rule{1.167pt}{0.108pt}}
\multiput(1366.00,496.17)(5.579,-3.000){2}{\rule{0.583pt}{0.400pt}}
\multiput(1374.00,492.95)(1.579,-0.447){3}{\rule{1.167pt}{0.108pt}}
\multiput(1374.00,493.17)(5.579,-3.000){2}{\rule{0.583pt}{0.400pt}}
\put(1382,489.17){\rule{1.700pt}{0.400pt}}
\multiput(1382.00,490.17)(4.472,-2.000){2}{\rule{0.850pt}{0.400pt}}
\multiput(1390.00,487.95)(1.355,-0.447){3}{\rule{1.033pt}{0.108pt}}
\multiput(1390.00,488.17)(4.855,-3.000){2}{\rule{0.517pt}{0.400pt}}
\multiput(1397.00,484.95)(1.579,-0.447){3}{\rule{1.167pt}{0.108pt}}
\multiput(1397.00,485.17)(5.579,-3.000){2}{\rule{0.583pt}{0.400pt}}
\multiput(1405.00,481.95)(1.579,-0.447){3}{\rule{1.167pt}{0.108pt}}
\multiput(1405.00,482.17)(5.579,-3.000){2}{\rule{0.583pt}{0.400pt}}
\multiput(1413.00,478.95)(1.579,-0.447){3}{\rule{1.167pt}{0.108pt}}
\multiput(1413.00,479.17)(5.579,-3.000){2}{\rule{0.583pt}{0.400pt}}
\multiput(1421.00,475.95)(1.355,-0.447){3}{\rule{1.033pt}{0.108pt}}
\multiput(1421.00,476.17)(4.855,-3.000){2}{\rule{0.517pt}{0.400pt}}
\multiput(1428.00,472.95)(1.579,-0.447){3}{\rule{1.167pt}{0.108pt}}
\multiput(1428.00,473.17)(5.579,-3.000){2}{\rule{0.583pt}{0.400pt}}
\end{picture}